\documentclass[conference]{IEEEtran}
\IEEEoverridecommandlockouts

\usepackage{tikz}
\usepackage{textcomp}
\usepackage{hyperref}
\usepackage{lipsum}

\usepackage{cite}
\usepackage{amsmath,amssymb,amsfonts}
\usepackage{algorithmic}
\usepackage{graphicx,epstopdf}
\usepackage{textcomp}
\usepackage{xcolor}
\usepackage{todonotes}
\def\BibTeX{{\rm B\kern-.05em{\sc i\kern-.025em b}\kern-.08em
    T\kern-.1667em\lower.7ex\hbox{E}\kern-.125emX}}
    
\newcommand\copyrighttext{
  \footnotesize \textcopyright 2020 IEEE. Personal use of this material is permitted. Permission from IEEE must be obtained for all other uses, in any current or future media, including reprinting/republishing this material for advertising or promotional purposes, creating new collective works, for resale or redistribution to servers or lists, or reuse of any copyrighted component of this work in other works.
  }
  
\newcommand\copyrightnotice{
\begin{tikzpicture}[remember picture,overlay]
\node[anchor=south,yshift=10pt] at (current page.south) {\fbox{\parbox{\dimexpr\textwidth-\fboxsep-\fboxrule\relax}{\copyrighttext}}};
\end{tikzpicture}
}    
    
\begin{document}

\title{Levels of Coupling in Dyadic Interaction: \\

An Analysis of Neural and Behavioral Complexity
}
\author{\IEEEauthorblockN{Georgina Montserrat Reséndiz-Benhumea}
\IEEEauthorblockA{\textit{Embodied Cognitive Science Unit} \\
\textit{Okinawa Institute of Science and}\\
\textit{Technology Graduate University}\\
Okinawa, Japan \\
\textit{Institute for Applied Mathematics}\\
\textit{and Systems Research}\\
\textit{National Autonomous University of Mexico}\\
Mexico City, Mexico \\
georginamontserrat.resendizbenhumea@oist.jp} 
\and
\IEEEauthorblockN{Ekaterina Sangati}
\IEEEauthorblockA{\textit{Embodied Cognitive Science Unit} \\
\textit{Okinawa Institute of Science and}\\
\textit{Technology Graduate University}\\
Okinawa, Japan \\
ekaterina.sangati@oist.jp}
\and
\IEEEauthorblockN{Tom Froese}
\IEEEauthorblockA{\textit{Embodied Cognitive Science Unit} \\
\textit{Okinawa Institute of Science and}\\ \textit{Technology Graduate University}\\
Okinawa, Japan \\
tom.froese@oist.jp}
}

\maketitle
\copyrightnotice

\IEEEpubidadjcol

\begin{abstract}
From an enactive approach, some previous studies have demonstrated that social interaction plays a fundamental role in the dynamics of neural and behavioral complexity of embodied agents. In particular, it has been shown that agents with a limited internal structure (2-neuron brains) that evolve in interaction can overcome this limitation and exhibit chaotic neural activity, typically associated with more complex dynamical systems (at least 3-dimensional). In the present paper we make two contributions to this line of work. First, we propose a conceptual distinction in levels of coupling between agents that could have an effect on neural and behavioral complexity. Second, we test the generalizability of previous results by testing agents with richer internal structure and evolving them in a richer, yet non-social, environment. We demonstrate that such agents can achieve levels of complexity comparable to agents that evolve in interactive settings. We discuss the significance of this result for the study of interaction.
\end{abstract}

\begin{IEEEkeywords}
agent-based modeling, social interaction, coupling, neural entropy, evolutionary robotics, continuous-time recurrent neural network, minimal cognition, dyad.
\end{IEEEkeywords}

\section{Introduction}

Recent years have seen an increase in efforts to understand the role of social interaction in social cognition from an embodied perspective \cite{DeJaegher-DiPaolo-2007, DeJaegher-DiPaolo-2008, DeJaegher-DiPaolo-2010, DiPaolo-2000, Froese-Gershenson-Rosenblueth-2013, Froese-2018, Candadai-et-al-2019}. It has been argued that rather than being merely an outcome of the dynamics of individual cognitive agents, social interaction can itself constitute cognition and have an effect on the individuals that partake in it\footnote{This is admittedly a contentious claim, cf. \cite{Herschbach-2012}.}. According to one specific embodied cognition account – enactivism – social interaction is defined as an active co-regulated coupling between two or more autonomous agents, where their role of interactors co-emerges with the interaction itself and their individual cognitive capacities can be reduced or augmented \cite{DeJaegher-DiPaolo-2007, DeJaegher-DiPaolo-2008, DeJaegher-DiPaolo-2010}. In this paper we focus on how different levels of coupling can influence the agent's neural and behavioral complexity.

The types of couplings distinguished in this and previous work are inspired by a well-known experiment from developmental psychology: the “double TV monitor” paradigm \cite{Murray-Trevarthen-1985}. In this experiment, 2-month-old infants interact with their mothers through a live video link. When the live video is replaced with a recorded replay of the previous actions of the mother, the infants become distressed, distracted and upset, suggesting that the reciprocity of the interaction makes a difference. That is, passive social input that is not sensitive to one’s own response is not sufficient for a positive social experience. Simulation studies described here show that it might also not be sufficient for individual cognition.

Based on the previous approaches, we propose to distinguish the following levels of coupling in dyadic interaction:
\begin{itemize}
\item 2-way or bidirectional coupling: Both agents are mutually interacting, e.g. normal interaction as a mother playing with her infant. Other examples are conversations, dancing, collaborative work, etc. \cite{DeJaegher-DiPaolo-2010}.
\item 1-way or unidirectional coupling: Active agent is in the presence of a non-interactive agent, which is showing pre-recorded behavior, e.g. “double TV monitor” experiment.
\item 0-way or no-way coupling: Active agent is not in the presence of a social partner, i.e. is alone.
\end{itemize}

This distinction is not meant to be exhaustive and future adjustments might be required. For instance, a case in which one agent is fully interactive while the other is present but staying still might fall somewhere between a 0-way and a 1-way coupling. Additionally, even within the isolated condition, one could distinguish different ways in which the agent can couple to its physical environment – bidirectionally in a full sensorimotor loop \cite{ORegan-2001} or one-directionally, whereby the agent is passively receiving stimulation from the environment. However, this work focuses on the social interaction scenario and we think it is still useful to think of levels of coupling within this domain and how they might have distinct effects on individual cognitive capacities.

\subsection{Previous work}
Evolutionary robotics (ER) has been used as a scientific tool to study minimal models of cognition. It is a methodology especially suited for embodied cognition because it allows for developing integrated sensorimotor systems which act in close coupling with their environments, i.e. ER takes into account both embodiment and situatedness in how they contribute to the solution of particular cognitive tasks \cite{Harvey-et-al-2005}. The Candadai et al. (2019) model \cite{Candadai-et-al-2019} is a minimal model of cognition in interaction based on ER. This model demonstrates as a proof of concept that social interaction transforms the neural and behavioral dynamics of embodied agents more than what is achievable by agents operating in isolation. In this model, pairs of agents were evolved in a 2-dimensional environment to maximize their individual neural entropy\footnote{It is important to emphasize here that maximizing this measure was not meant to achieve any particular adaptive outcomes. Rather, the point is to establish whether 2D brains can be evolved to become more complex and whether interaction is an enabling condition for this.}. When the agents were evolved together (interaction condition) they exhibited mutually coordinated behavior and higher individual neural entropy compared to when they were evolved alone (isolation condition). Furthermore, when agents that evolved in the interaction condition were tested in the presence of a "ghost" non-interactive partner (ghost condition), they exhibited a loss in neural and behavioral complexity. This condition is analogous to a “double TV monitor experiment” in which infants used to normal 2-way interaction mode are suddenly placed in a 1-way interaction mode.

In our previous work \cite{Resendiz-Benhumea-Froese-2020}, we have replicated the Candadai et al. (2019) model \cite{Candadai-et-al-2019} with a less constrained parameter space of the agents, leading to a broader diversity in the agent genotypes but achieving similar results in terms of neural complexity. We also provided a state-space analysis of the evolved agents’ neural network activity, which showed a single fixed-point attractor. This is consistent with previous ER work that shows that CTRNN controllers are often able to display rich dynamics despite having a single attractor because the attractor landscape is constantly shifting in the agent’s interaction with the environment \cite{Beer-2000, Buckley-2008}.

\subsection{Current work}
The Candadai et al. (2019) model \cite{Candadai-et-al-2019} demonstrated that 2-way coupling leads to higher neural complexity than 1-way coupling and 0-way coupling. However, the specific implementation adopted is open to two central criticisms. 

First, it is known that continuous dynamical systems can only exhibit chaotic behavior when they have more than 2 dimensions \cite{Coddington-Levinson-1955}. This is indeed an assumption of the Candadai et al. (2019) model \cite{Candadai-et-al-2019}: the fact that the agents’ 2D brains exhibit chaotic activity is remarkable because it means that interaction can overcome the inherent limitations of their brains. However, a corollary of this is that if agents had brains with more dimensions, they could generate neural and behavioral complexity without interaction. If so, this would limit the result to very constrained systems and tell us nothing about cognition in multi-dimensional actual brains.

Second, it could be argued that a comparison between agents evolved in isolation and agents evolved in interaction is not a strong comparison that would justify far-reaching claims about the role of interaction in cognitive complexity. The isolated agents might be less complex not only because they don’t interact but because they evolve in an overall poorer environment from which they receive no input. It could be that if the environment contained some stimulation, even if not social, the agents could leverage it to create more neural complexity.

\begin{figure}[htbp]
\centering
\includegraphics[width=1.9in]{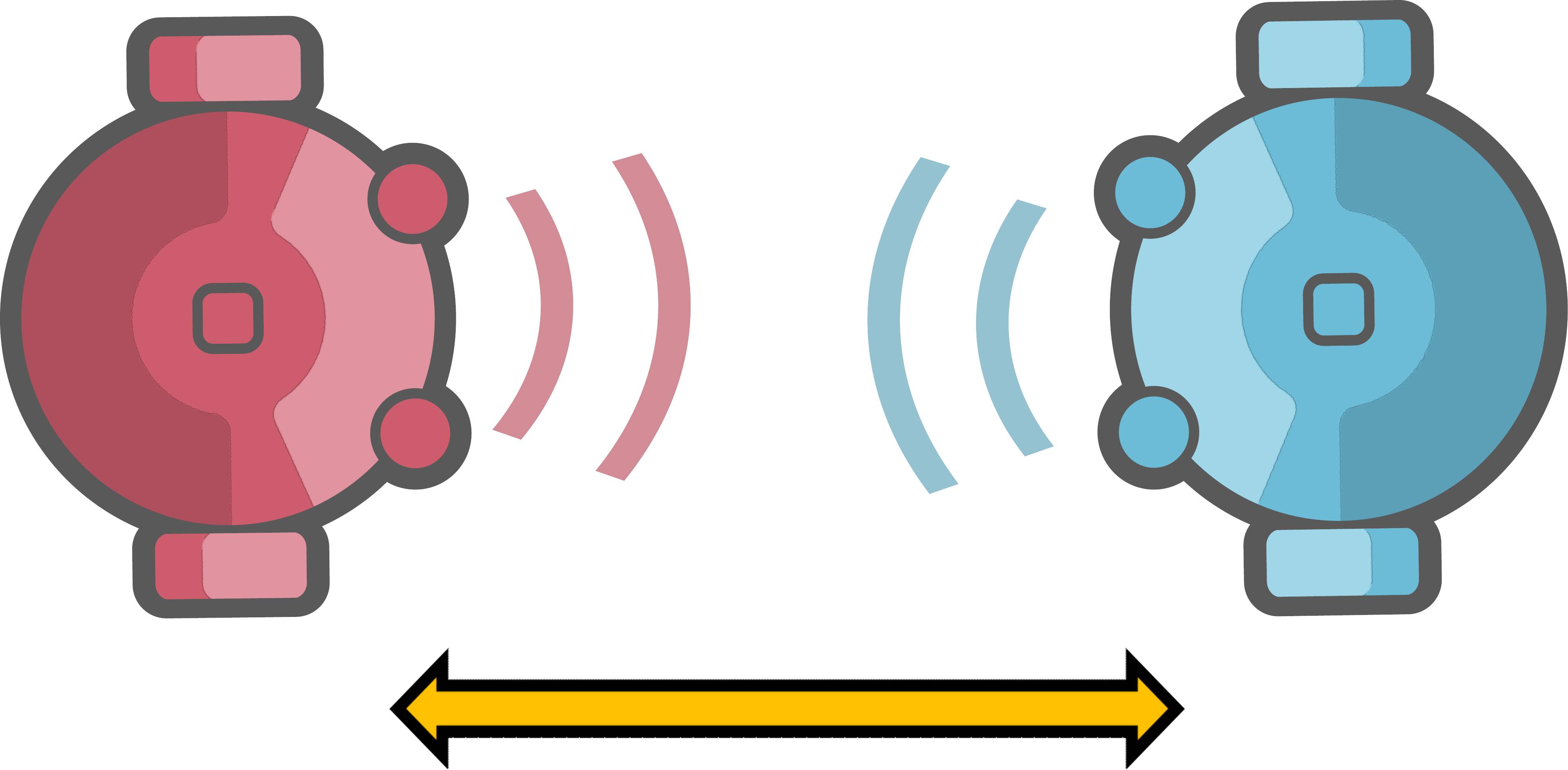}
\caption{Two-way coupling: Fully interactive condition, agents interact with each other by sending and receiving acoustic signals.}
\label{fig_1}
\end{figure}

\begin{figure}[htbp]
\centering
\includegraphics[width=1.9in]{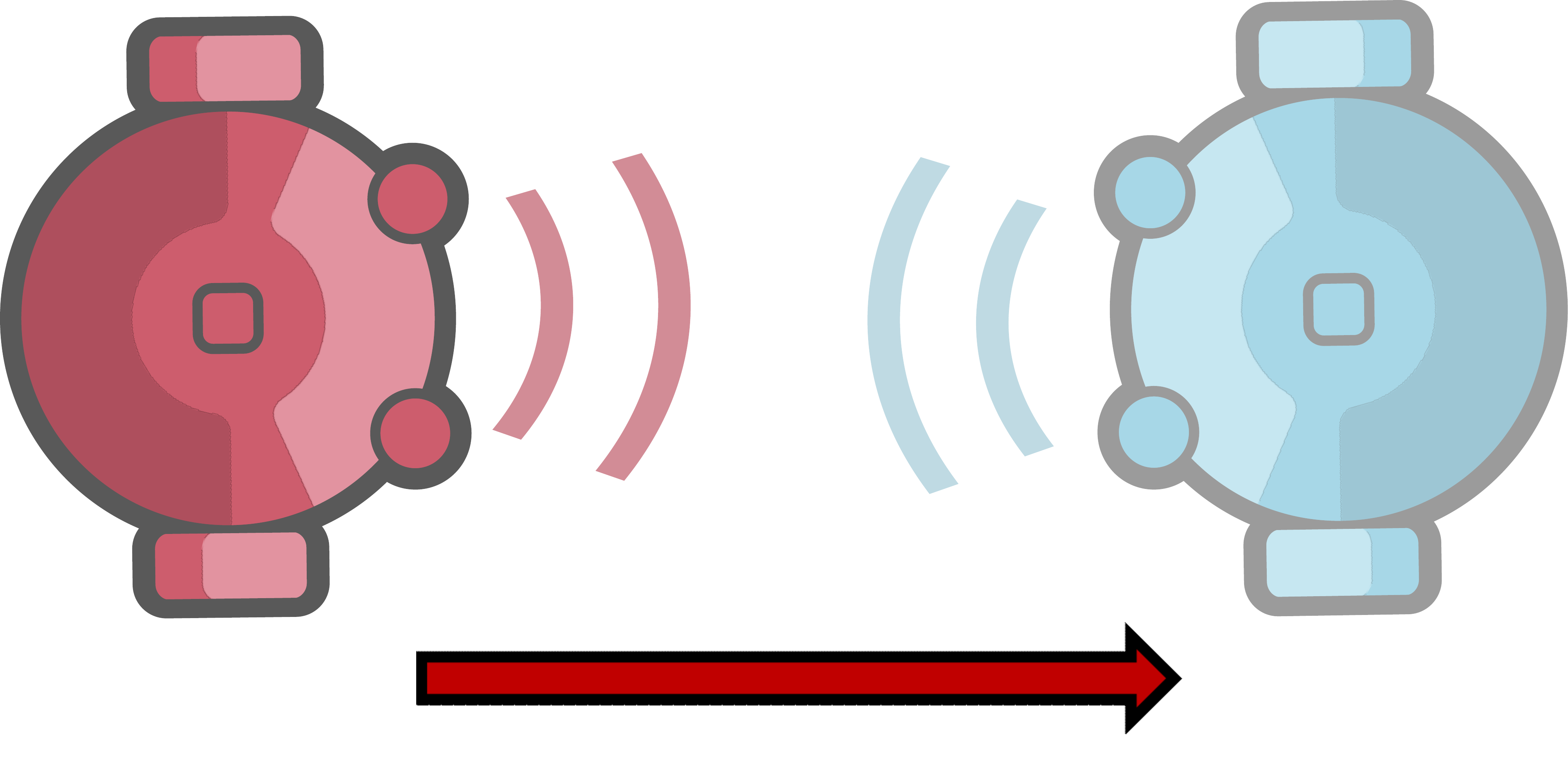}
\caption{One-way coupling: “Live” agent (red) tested or evolved with a “ghost”
agent (blue). Live agent is sending and receiving acoustic signals. Ghost agent
is playing back pre-recorded behavior of previous trials and so sending pre-recorded signals but not receiving anything.}
\label{fig_2}
\end{figure}

\begin{figure}[htbp]
\centering
\includegraphics[width=0.59in]{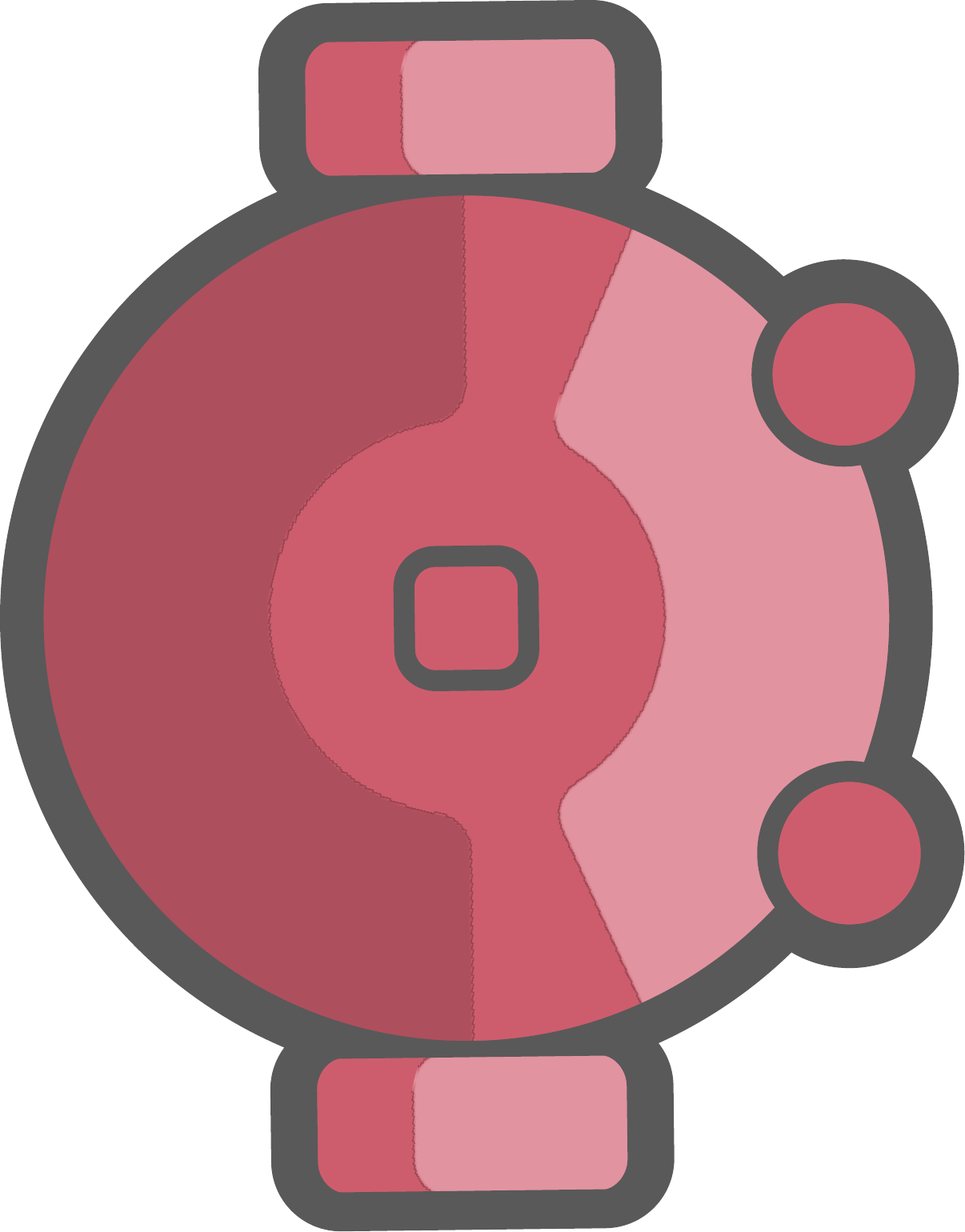}
\caption{Zero-way coupling. The “isolated” agent (red agent) is evolved on its
own, therefore, the agent neither receives any input from the environment nor
interacts with other agents.}
\label{fig_3}
\end{figure}

In this study, we address these two limitations and investigate whether the Candadai et al. (2019) model \cite{Candadai-et-al-2019} results still hold when we 1) include one more neuron in the internal layer of the neural architecture of the agents, i.e. we construct a 3-neuron model instead of 2-neuron model and 2) we create a new condition in which we evolve agents with a ghost agent, thereby providing them with a source of non-interactive stimulation. As a result, we implement 4 conditions based on 3 levels of coupling distinguished above, as illustrated in Fig.~\ref{fig_1},~\ref{fig_2},~\ref{fig_3}:
\begin{itemize}
\item \textbf{Interactive condition (2-way coupling):} pairs of agents are evolved in bidirectional interaction.
\item \textbf{Ghost-test condition (1-way coupling):} pairs of agents are evolved in bidirectional interaction but tested with a ghost at the end.
\item \textbf{Ghost-evolution condition (1-way coupling):} an active agent is evolved in the presence of a sufficiently complex ghost.
\item \textbf{Isolated condition (0-way coupling):} agents are evolved in isolation.
\end{itemize}

In all conditions agents are evolved to maximize their neural complexity operationalized, for simplicity, as Shannon entropy of neural outputs. However, the maximization of predictive information (PI, \cite{Bialek-1999}), which has been successfully applied in previous self-organizing robotic systems where they exhibited a wider behavioral spectrum \cite{Martius-Der-Ay-2013}, is considered as an alternative approach for future work.

We then compare the level of neural and behavioral complexity in the last generation across all conditions. We find that despite having more powerful brains, agents that evolve in isolation exhibit lower neural and behavioral complexity. However, agents that evolve in the presence of the ghost exhibit an interesting divergence between the complexity of their neural output and behavior: their neural complexity is comparable to that of agents that evolve in interaction while their behavioral complexity is lower. We also investigate the quality of interaction between social conditions and conclude by placing our results in a broader theoretical context of embodied cognition.

\section{Methods}
\subsection{Model Design}
Our model is a replication of the Candadai et al. (2019) model \cite{Candadai-et-al-2019}, except for the increased number of neurons in the internal layer of the neural architecture: three neurons instead of two. 
\subsubsection{Agents and environment} 
Agent bodies are circular, with a radius of 4 units. Each of them has two acoustic sensors symmetrically placed in the front of the agent at a 45$^{\circ}$ angle to the central axis; an acoustic emitter positioned in the center of the body; and two motors located in the right and left sides of the agent to allow movement in a 2-dimensional empty environment, which is an unlimited arena. Collisions are modeled as point elastic, which means, no change in the agents’ angular velocity (i.e. no friction between them) and zero net effect on their velocity vectors (i.e. energy of the complete system is conserved). This is achieved by exchanging the agents’ velocity vectors, which causes them to bounce off each other without loss of energy \cite{DiPaolo-2000}. Each agent can emit and sense acoustic signals. The strength of the acoustic signal experiences two kinds of attenuation:
\paragraph{Attenuation due to distance} The maximum strength of the emitted signal is exhibited from the source to a distance equal to 2R between the center of the bodies of the agents. The intensity of the signal obtained by each sensor is then calculated by applying the inverse square law using the distance between the sensor and the source.
\paragraph{Attenuation due to “self-shadowing” mechanism}  This attenuation occurs when the emitted signal passes within the body of the sensing agent (as the intensity of the acoustic signal is weakened due to the agent's own embodiment). It is modeled as a scaling factor over the sensory inputs in a range from 0.1 (when the sensors of the sensing agent are diametrically opposite from the source) to 1 (when the sensing agent is facing the source). The equations to calculate the distance that the emitted signal travels within the body of the sensing agent, i.e. the shielded distance (\(D_{sh}\)), are available in the Supplementary Material of the Candadai et al. (2019) model \cite{Candadai-et-al-2019}.

According to the previous points, the sensory input for each sensor of an agent is calculated using the distance between the sensor and the source (applying the inverse square law), and it is then multiplied by the “self-shadowing” attenuating factor which goes linearly from 1 (when \(D_{sh}=0\)) to 0.1 (when \(D_{sh}=2R\)).

\subsubsection{Neural architecture} 
\begin{figure}[tbp]
\centering
\includegraphics[width=2.7in]{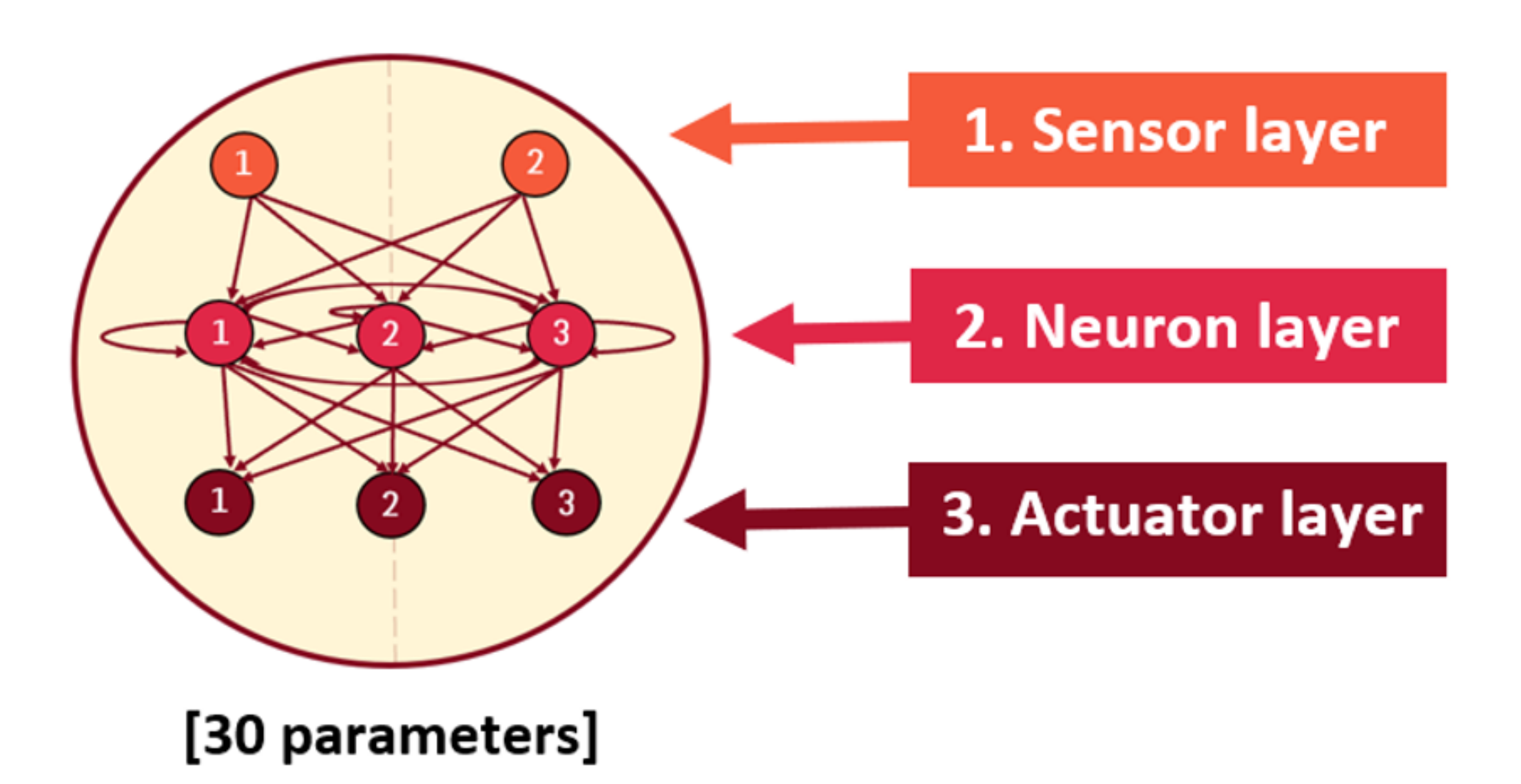}
\caption{Neural architecture of the 3-neuron model based on the Candadai et al.
(2019) \cite{Candadai-et-al-2019} model, where the number of neurons in increased to three. In this
approach, the two sensor nodes share common gain and bias; the three neurons
share common time-constant and bias; and, the three actuator nodes share
common gain and bias.}
\label{fig_4}

\end{figure}

The neural architecture of each agent consists of three layers, in our previous work we called them: sensor layer, neuron layer and actuator layer. The main difference from our current model and the Candadai et al. (2019) model \cite{Candadai-et-al-2019} is presented in the neuron layer, where instead of two neurons, we use three neurons, as shown in Fig.~\ref{fig_4}

\paragraph{Sensor layer} In this layer, there are two sensor nodes with a sigmoidal activation function, whose output is given by:

\begin{equation}
o_{s}=g_{s}\sigma(I_{s}+\theta_{s})\label{eq_1}
\end{equation}

where \(gs\) is the sensory gain, \(\sigma(x)=1/(1+e^{-x})\) is the sigmoidal activation function, \(I_{s}\) is the sensory input and \(\theta_{s}\) is the bias. In this layer, both sensor nodes share common gain and bias. 

\paragraph{Neuron layer} This layer is modeled as a continuous-time recurrent neural network (CTRNN) \cite{Beer-1995}. In contrast to the Candadai et al. (2019) model \cite{Candadai-et-al-2019}, this layer now consists of three fully recurrently connected neurons, which corresponds to a 3-dimensional dynamical system. Each neuron’s activity is governed by the following state equation: 

\begin{equation}
    \tau_{i} \frac{dy_{i}}{dt}=-y_{i}+\sum_{j=1}^{N} w_{ij}\sigma(y_{j}+\theta_{j})+\sum_{s=1}^{2} w_{is}o_{s}
\end{equation}

where \(dy_{i}/d_{t}\) refers to the rate of change of internal state \(y_{i}\) of neuron \(i\) based on a time constant \(\tau_{i}\). The rate of change \(dy_{i}/d_{t}\) depends on the current state of the neuron, the weighted sum of outputs from other internal neurons and the total external input from the sensors. The output of each neuron based on its internal state is given by a sigmoid activation function \(\sigma(y_{i}+\theta_{j})\) where \(\theta_{j}\) refers to the neuron's bias. In this implementation, the three neurons share same time-constant and bias.

\paragraph{Actuator layer} The three internal neurons feed into the actuator layer, where the input to each actuator node is a weighted sum of the outputs of the neuron. The actuator layer contains three actuator nodes, two corresponding to the motors and one corresponding to the acoustic signal emitter. All of them are sigmoidal units with a gain and bias (but no internal state) such that the output of the actuator node \(i\), \(m_{i}\), is given by:

\begin{equation}
    m_{i}=g_{m}\sigma \left( \sum_{n=1}^{N} w_{ni}*o_{n} + \theta_{i} \right)
\end{equation}

where \(o_{n}\) is the output of the neuron, that are weighted by \(w_{ni}\), \(\theta_{i}\) is the bias and \(g_{m}\) is the gain. In this layer, the three actuator nodes share common gain and bias.

Locomotion is managed by the effective control of the two motors. Net linear velocity is given by the average of their corresponding outputs and angular velocity which rotates the agent is given by their difference divided by the radius of the agent.

\subsubsection{Evolutionary optimization}
\paragraph{Fitness function: Neural entropy}
The fitness function for the evolutionary algorithm is the Shannon entropy of neural outputs, i.e. neural entropy, which has been used as a proxy of cognitive complexity. This function does not optimize any particular task. The agents are initialized in a random position in the environment (see below) and allowed to move around over 4 trials, during which their neural activity is recorded. The neural complexity is measured as the Shannon entropy in the three-dimensional time series from the outputs of the three neurons, which are bounded in the range from 0 to 1. The output space is binned with 100 bins along each of the three dimensions, i.e. one million bins in total. Then, a 3-dimensional histogram is created using all the binning data points acquired during the 4 trials. Thus, the Shannon entropy \(H\) of the neural time series is given by:

\begin{equation}
    H=\sum_{i=1}^{100} \sum_{j=1}^{100} \sum_{k=1}^{100} -p_{ijk} log(p_{ijk})
\end{equation}

where the probability of the neural activity in a specific bin \([i,j]\), \(p_{ij}\), is given by the number of data points in that bin divided by the total number of data points. The neural entropy is then normalized to be in the range from 0 to 1 by dividing by the maximum neural entropy, i.e. \(log(100*100*100)\), when all bins are uniformly populated. Therefore, the normalized neural entropy is given by: 

\begin{equation}
    \hat{H} = H/log(100*100*100)
\end{equation}

\paragraph{Genetic algorithm}
We used a real-valued genetic algorithm to optimize the parameters of the neural controllers, such as weights, gains, biases and time-constants in order to maximize the neural entropy of the agents. Each agent had 30 parameters, i.e. for N agents, the genotype consisted of 30N parameters that were initially encoded in the range \([-1, 1]\). When building the agents from each genotype to perform the 4 trials, these parameters were scaled appropriately, following the same parameter ranges as in the Candadai et al. (2019) model \cite{Candadai-et-al-2019} such that: for sensor and actuator nodes, the gains were scaled in the range \([1, 5]\) and the biases were scaled in the range \([-3, 3]\); for neuron nodes, the time-constants were scaled in the range \([1, 2]\) and the biases were scaled in the range \([-3, 3]\); all weights were scaled in the range \([-8, 8]\). 

The performance evaluation of the agents was obtained according to each of the 4 conditions: interactive condition, ghost-test condition, ghost-evolution condition and isolated condition. The experimental setup for each of the 4 conditions is described in detail in the Experiments section. 

For all conditions, after the performance evaluation, we generated a new population by, first, keeping an elite population of the top 4\% solutions as it is and, second, by mutating and crossing over this elite fraction to get the remainder of the solutions. Mutation was obtained by adding a zero-mean Gaussian mutation noise with variance 0.1 to the solutions and, then, crossover was obtained by swapping each parameter between a pair of solutions with a probability of 0.1.

\subsection{Experiments}
Here, we describe the implementation of the 4 conditions introduced in the Introduction:

\subsubsection{Condition 1: Interactive condition (2-way coupling)}
In this condition, we performed 100 independent runs using 96 pairs of agents that were able to interact with each other. Each pair’s agent parameters were encoded in a single genotype subjected to evolutionary search. Initially, the agents were placed at 20 units from each other. For each trial, their relative angle was modified as \([0, \pi/2, \pi, 3\pi/2]\), respectively, where both agents' heading direction was set to the right. The population was evolved for up to 2000 generations to maximize the neural entropy of both agents. 

\subsubsection{Condition 2: Ghost-test condition (1-way coupling)}
In this condition, we selected the best pair of agents of the best 10 runs previously obtained in the fully interactive scenario and, then, tested them under a “ghost” condition. Red agent was the “live” agent and blue agent was the “ghost” agent. The “live” agent was able to interact with the “ghost” partner, while, the “ghost” agent was just playing back pre-recorded behavior from the previous trials in a fully interactive scenario. The “live” agent was initially positioned at a different angle (randomly selected from \([0, \pi/2, \pi, 3\pi/2]\) but different to the one it was chosen when the "live" agent was in 2-way interaction) from the “ghost” agent to avoid repeating the behavior of those previous trials. The initial distance between both agents was 20 units. We conducted 4 trials for each pair of agents, as in the fully interactive condition, and measured the normalized neural entropy of the “live” agent. 

\subsubsection{Condition 3: Ghost-evolution condition (1-way coupling)}
In this condition, we selected the best blue agent of the best run in fully interactive scenario and used it as the non-interactive or “ghost” agent. We performed 10 independent runs, where only the interactive or “live” agent was evolved in the presence of the “ghost” partner, which was just playing back the pre-recorded behavior in fully interactive scenario (different for the 4 trials). The same playback was used for all runs, all agents and all generations, in order to evolve the agents to respond to a specific set of conditions and see the effects on neural and behavioral complexity through each generation. The population in each run consisted of 96 individuals, where each individual was encoding the parameters of only the “live” agent. For each trial, the "live" agent was placed 20 units from the "ghost" agent and their relative angle was modified as \([0, \pi/2, \pi, 3\pi/2]\), respectively. The population was evolved for up to 2000 generations to maximize the interaction entropy of “live” agents. Evaluation was performed the same way as for Condition 1 with fitness derived only from the “live” agent’s neural entropy.

\subsubsection{Condition 4: Isolated condition (0-way coupling)}
In this condition, we performed 10 independent runs using isolated agents that were not receiving any input and were evolved on their own to maximize their isolation entropy. Red agent was referred as the “isolated” agent. The population in each run consisted of 96 individuals, where each individual was encoding the parameters for only one agent (“isolated” agent). The population was evolved up to 2000 generations. 

In all conditions, the agents' initial heading direction was set to the right.
    
\section{Results}
\begin{figure*}[tbp]
\centering
\includegraphics[width=7.0in]{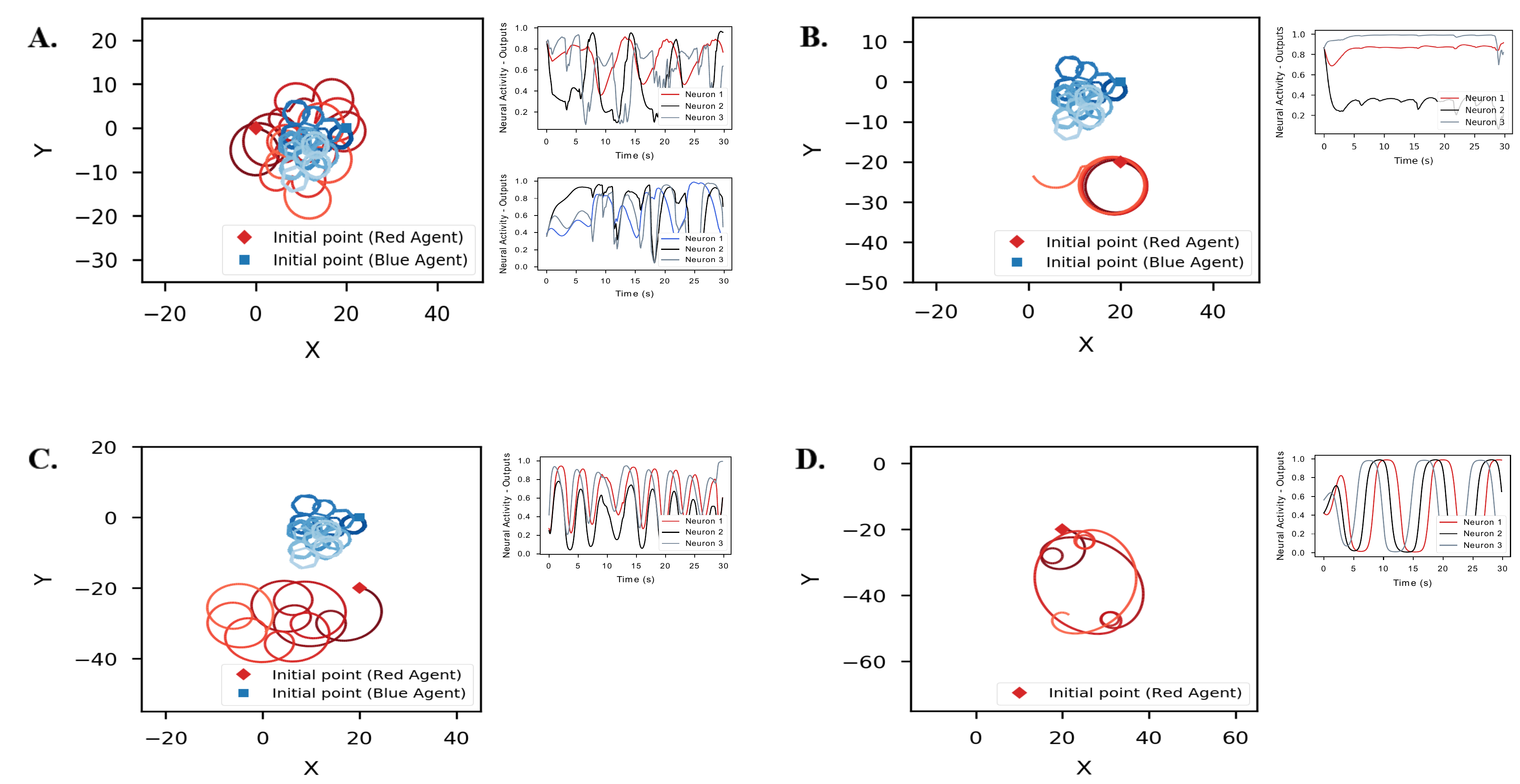}
\caption{Example plots of movement trajectories and neural activity across 4 conditions. Movement trajectories in red are of a live agent, movement
trajectories in blue are of a live agent in Figure A and of ghost replay in figures B and C. Neural activity plots show output of 3 internal neurons of the live agents.}
\label{fig_5}
\end{figure*}

 Fig.~\ref{fig_5} shows example trajectories and neural activation of the best agents in 4 conditions from the best runs, respectively. 
Complex behavior and neural activity can be clearly seen on the fully interactive condition. This complexity seems to be lost in the ghost-test condition, in line with Candadai et al. (2019) model \cite{Candadai-et-al-2019} results. The agent that exhibits complex movement trajectory when interacting with a live partner, starts to literally run in circles when the partner is non-responsive. This is the case even though the agents in the current experiment have a more complex 3-neuron brain that could in principle exhibit chaotic activity. In the isolated condition, the agent shows highly regular behavior and an oscillatory pattern of neural activity, again, in line with the results of the original model and despite a more complex brain. The most interesting for our purposes ghost-evolution condition, in which the agents are evolved in the presence of a ghost partner displays something that seems as an intermediate level of behavioral and neural complexity between the fully interactive and the isolated case.

In order to go beyond intuitions and understand the general pattern of differences between conditions, we run 4 statistical tests comparing the best agents' neural activity and movement trajectories. 

\begin{figure}[htbp]
\centering
\includegraphics[width=3.2in]{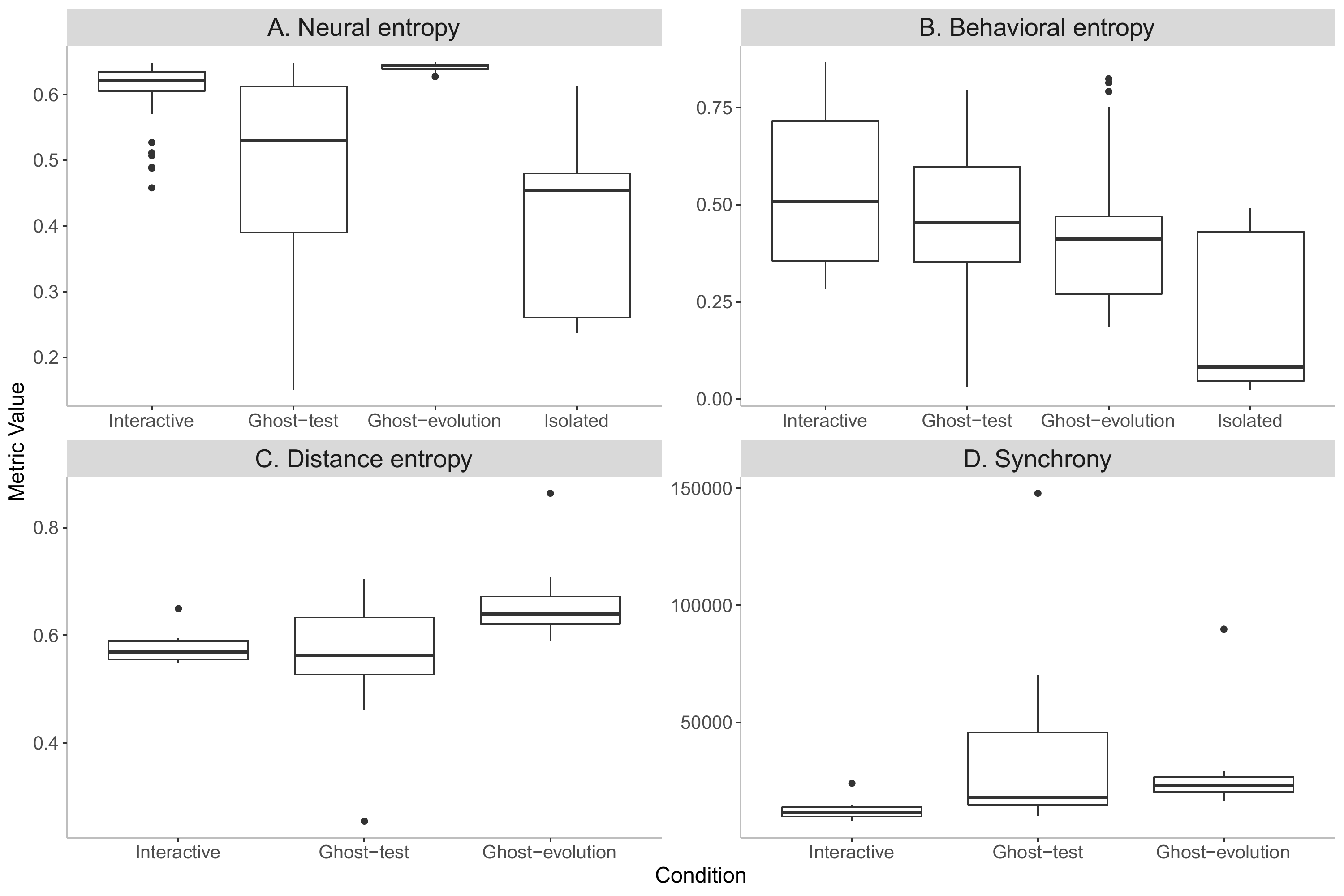}
\caption{Individual and interaction measures across experimental conditions.
Figures A and B show neural and behavioral entropy of live agents. C:
distance entropy between agents. D: DTW-estimated distance between
agents (lower distance means more synchrony).}
\label{fig_6}
\end{figure}

Specifically, in the first test we compared the average of the means of neural entropy of 100 agent pairs in interactive condition against the means of the best live agents in other conditions. Given unequal sample size and significantly different variances between conditions, \(F(3, 127)= 21.24\), \(p<.001\), we performed a non-parametric Kruskal-Wallis test, which showed a significant difference in neural entropy between conditions, \(H(3)= 41.7\), \(p<.001\). Focused comparisons of the mean ranks between groups showed that agents in isolated condition had significantly lower neural entropy compared to interactive and ghost-evolution conditions, as expected. However, somewhat surprisingly, the entropy in ghost-evolution condition was significantly higher than in the interactive or ghost-test conditions. This trend can also be seen in Fig.~\ref{fig_6}A.

Next, we obtained a measure of behavioral complexity of the live agents in all conditions. We recorded heading direction angles at each time point of agent trajectories, which resulted in 1D time series. We then computed sample entropy for each such time series. Since there does not seem to be a universally agreed upon measure of behavioral trajectory complexity, we used sample entropy as a measure that has been shown to be an appropriate index of complexity for biological time series more broadly \cite{Richman-2000}. We have informally validated this measure by checking that it reliably distinguishes between fully regular circular and more complex movement patterns\footnote{We originally tried to estimate behavioral complexity using image entropy calculated on generated plots of trajectories. However, this measure delivered inconsistent results, most likely because it does not consider the temporally ordered nature of movement coordinates and instead takes into account only their spatial dispersion.}. The data subjected to statistical analysis were time series sample entropy for all trials of the 10 best runs in each condition. In the interactive condition, only one of live agents was considered. Levene’s test for equality of variances was not significant and there was a significant effect of condition on the level of behavioral entropy, \(F(3,153)= 20.98\), \(p<.001\). Post-hoc Bonferroni-corrected tests showed that isolated agents’ entropy was significantly lower than entropy in all other conditions \((p<.001)\) but also ghost-evolution condition entropy was significantly lower than entropy in the interactive condition (see Fig.~\ref{fig_6}B).

As the difference in neural and behavioral complexity in the ghost-evolution condition compared to interactive condition was in opposite directions (higher and lower respectively), we run two further tests to investigate the type of interaction between the agents in 3 conditions with coupling (thus, excluding the isolated condition). In the first test we have computed the entropy of the distance between the agents while in the second their synchrony. Distance entropy was measured with a binned Shannon entropy approach adopted in the original Candadai et al. (2019) paper \cite{Candadai-et-al-2019} . Synchrony was estimated with a Dynamic Time Warping metric applied to 2-dimensional time series representing the xy-coordinates of the two agents in all trials. Only the first measure showed a significant overall effect of condition, \(F(2, 27)=4.27\), \(p<.05\) and a significant post-hoc pairwise comparison between Ghost-evolution and Ghost-test conditions. Fig.~\ref{fig_6}C and Fig.~\ref{fig_6}D show the overall trends for these measures.

\section{Discussion and Conclusions}
In this work, we have extended the Candadai et al. (2019) model \cite{Candadai-et-al-2019} in order to address some of its limitations. By including one more neuron in the neuron layer, i.e. three neurons instead of two, we have found that the results in terms of neural and behavioral complexity are similar to those in the original configuration. In particular, the agents’ neural entropy is still higher in the interactive condition than in the ghost-test or isolated conditions. Additionally, we have quantitatively assessed behavioral complexity in all conditions and this measure was found to be also lower in the isolated condition. This means, perhaps unsurprisingly, that a powerful brain operating in isolation, without any input is not able to achieve high levels of neural or behavioral complexity. 

Our second generalization test delivered mixed insights. On the one hand, by observing example plots of neural activation (Fig. 5), it can be suggested that in the interactive condition, the neural activity of the three neurons exhibits more chaotic activity than in the other conditions, including the more stringent ghost-evolution condition. On the other hand, this suggestion is not borne out by the statistical test that shows that neural complexity in this condition is higher than in the interactive condition. This would mean that richer environment that provides constant but complex input to the agent can compensate the relative poverty of the 1-way coupling and lack of contingent response of the interaction partner. At the same time, despite a higher neural complexity, behavioral complexity in this setting is lower. Thus, agents that are evolved to maximize their neural complexity in a rich environment end up behaving in more predictable and regular manner than agents that receive specifically social stimulation. 

Briefly, our findings suggests that the richness of the environment may compensate 1-way interaction in terms of neural complexity, however, this does not apply in terms of behavioral complexity. This suggests that the social world (i.e. real-time interaction between agents becoming the whole brain-body-environment-body-brain system \cite{Froese-2013}) allows for a greater repertoire of behaviors transforming our individual capacities \cite{DeJaegher-DiPaolo-2010}. 

Some possible limitations may be encountered in our model based on the specific metrics that we used as a measure of neural and behavioral complexity. Further work is needed to compare the results by implementing alternative methods, e.g. evolving agents to maximize predictive information (PI), using permutation entropy and applied it to raw xy-coordinates, etc. 

\section{Future Work}
In this paper, we have explored the neural and behavioral complexity of embodied agents using different levels of coupling in dyadic interaction. Future work will investigate how different modes of coupling can affect individual and interactive capacities of evolved agents. Specifically, it could be argued that an opportunity to interact with multiple partners could further enhance individual complexity. Alternatively, allowing agents to use different interactive modalities, such as distal and proximal coupling analogous to pheromone and saliva-based interactions in ants, could enrich their cognition. This will allow us to further understand how individual complexity can be generated by interaction.

\section*{Acknowledgment}
We are grateful to Alexey Yudin for sharing with us useful suggestions for the analysis of the model.

The authors thank the Scientific Computing \& Data Analysis Section (SCDA) of Research Support Division at Okinawa Institute of Science and Technology (OIST) for using their High-Performance Computing resources.

We acknowledge the help from Randall D. Beer by providing the Evolutionary Agents C++ software package v1.2.

\end{document}